\documentclass{article}

\usepackage{amsmath}
\usepackage{arxiv}
\usepackage[utf8]{inputenc} 
\usepackage[T1]{fontenc}    
\usepackage{hyperref}       
\usepackage{url}            
\usepackage{booktabs}       
\usepackage{amsfonts}       
\usepackage{nicefrac}       
\usepackage{microtype}      
\usepackage{mathtools}
\usepackage{url}
\usepackage{float}
\usepackage{placeins}
\usepackage{setspace}
\usepackage{adjustbox}
\usepackage{lineno}
\usepackage{natbib}
\usepackage{bm}
\usepackage{upgreek}
\usepackage{textcomp}
\usepackage{multirow}

\makeatletter
\newif\ifnobrackets
\renewcommand\@cite[2]{\ifnobrackets\else[\fi{#1\if@tempswa , #2\fi}\ifnobrackets\else]\fi\nobracketsfalse}

\makeatother

\title{Bayesian inference of 1D activity profiles from segmented gamma scanning of a heterogeneous radioactive waste drum}

\author{
   Eric Laloy\thanks{\normalsize{Corresponding Author, \texttt{eric.laloy@sckcen.be}}} \\
   \And
   Bart Rogiers \\
   \And
   An Bielen \\
   \And
   Sven Boden \\  
}

\begin{document}
\maketitle

\begin{abstract}
We present a Bayesian approach to probabilistically infer vertical activity profiles within a radioactive waste drum from segmented gamma scanning (SGS) measurements. Our approach resorts to Markov chain Monte Carlo (MCMC) sampling using the state-of-the-art Hamiltonian Monte Carlo (HMC) technique and accounts for two important sources of uncertainty: the measurement uncertainty and the uncertainty in the source distribution within the drum. In addition, our efficiency model simulates the contributions of all considered segments to each count measurement. Our approach is first demonstrated with a synthetic example, after which it is used to resolve the vertical activity distribution of 5 nuclides in a real waste package.
\end{abstract}


\section{Introduction}
\label{intro}

Segmented gamma scanning (SGS) is a longstanding but still actively researched technique to assess the activity of radioactive waste drums in a non-destructive way \citep[e.g.,][and references therein]{Bai2009,Krings-Mauerhofer2011,Patra-Agarwal2019,Frosio2020}. Classical SGS interpretation methods either sum up all gamma spectra to derive representative activities for the whole drum or treat the measured counts associated with each segment individually. Efficiency calibration can be performed in a classical way, placing a large amount of sealed radioactive reference sources in various drum matrices and simulate source homogeneity \citep[e.g.,][]{Marijuan2017}. Nowadays, Monte Carlo (MC) modeling techniques such as implemented in the commercially available ISOCS software \citep[ISOCS: In Situ Object Counting System,][]{ISOCS} are commonly used to determine detector's efficiency. Though offering significant geometrical flexibility, the ISOCS variants cannot always fully account for the spatial distribution of activities within the drum. Furthermore, the uncertainty in the estimated activities by radionuclide quantification techniques is often calculated in a relatively simple way, whether based on first-order Taylor expansion or MC error propagation, ignoring for instance drawbacks related to non-Gaussian distributions or non-linear expressions, or the omission of, e.g., systematic uncertainties like that of the efficiency \citep[e.g.,][]{Kirkpatrick2013}. Bayesian inference, also called Bayesian data inversion, is well suited to assess the uncertainty of quantities that result from the application of a mathematical model \citep[see the book by][for a comprehensive description of Bayesian data analysis]{Gelman2014}. In our opinion, Bayesian inference is particularly attractive for radiological characterization for the following reasons.

\begin{itemize}
	\item We are not simplifying the mathematical model, as is the case for first-order Taylor expansion, and can work with any type of distribution (this is the same for MC error propagation)
	
	\item We can account for prior information and expert opinion. Also MC error propagation allows this. The difference is however, that if there is information in the data, the Bayesian inference will update those prior beliefs into posterior beliefs, while in the MC error propagation, there is no way to do that. Also, for the Bayesian inference, we can express our prior beliefs on the quantities of interest (e.g. activities), while for the MC error propagation there is no way to do that.
	
	\item We do in principle not need to give special attention to performance characteristics like decision threshold, detection limit and minimum detectable activities (MDA). The Bayesian inference will, based on our prior beliefs, always provide the possible range of outcomes, even for cases where a classic workflow would have resulted in a reported MDA.
	
	\item With this approach it is straightforward to deal with situations for which multiple observations inform about the same quantities and/or processes. For instance, when one or more inferred radionuclide activities are associated with multiple energy peaks in gamma spectrometry. Here for a given nuclide, Bayesian analysis will provide a unique posterior activity distribution that is consistent with all the energy peaks. In contrast, this is something neither conventional first-order Taylor expansion nor MC uncertainty propagation can cope with. Indeed, in this case the uncertainty propagation would deliver separate activity distributions for each peak, with no straightforward way to combine them. Other situations of multiple observations informing about the same quantity arise when different radiological measurement techniques sensing the same radionuclide activities and/or relying on the same efficiency model are combined. 
\end{itemize}

The pioneer conference paper by \citet{Clement2018} introduced the use of Bayesian analysis in the radioactive waste characterization community, using a standard random walk Metropolis (RWM) Markov chain Monte Carlo (MCMC) algorithm \citep[see][for details about MCMC]{Gelman2014}. Recently, \citet{Carasco2021} presented a Bayesian framework to infer the mass of a given radionuclide contained in an horizontal drum slice from coupled gamma-ray spectrometry and tomographic scanning. In this work, we propose a fully Bayesian approach to infer the vertical distribution of the activities of the measured nuclides and to quantify the corresponding uncertainty. Our approach relies on state-of-the-art MCMC sampling using the Hamiltonian Monte Carlo (HMC) technique \citep[][]{Neal2011,Betancourt2018} and accounts for the net and background count uncertainties (i.e., Poisson statistics) together with the uncertainty in the detector's efficiencies caused by the uncertainty in the source distribution. Furthermore, our efficiency model accounts for the contributions of all segments to each count measurement given the assumption of a constant matrix density and corresponding gamma ray attenuation. To the best of our knowledge, this is the first application of Bayesian inference to estimation and uncertainty quantification of spatially-distributed activity from SGS measurements.

The remainder of this paper is organized as follows. Section \ref{methods} describes the considered waste drum, the used SGS apparatus and the measured data, together with the considered model and the MCMC-based Bayesian inversion. Section \ref{results} then details the results of our high-dimensional MCMC sampling for spatially-distributed radionuclides' quantification for both a synthetic and a real case, before section \ref{discussion} provides some discussion and outlines some future developments. Finally, section \ref{conclusion} summarizes our main findings and provides a short conclusion.

\section{Methods}
\label{methods}

\subsection{Segmented gamma scanning}
\label{sgs}

The non-destructive technique used in this study to identify and quantify the gamma-emitters present in a radioactive waste drum is segmented gamma scanning (SGS), which scans the drum slice by slice (segment), while the waste drum is rotating. We used the so-called 3AX SGS \citep{Buecherl1998}. The 3AX device can scan in the horizontal plane while the drum is rotating. Here, we only used the rotation capability and varied the height of the 3AX detector along the vertical axis of the drum. Also, we did not further subdivide the scanned segments in sectors. A 3AX SGS is made of three parts: the mechanics, the detector system and the data acquisition and processing unit. The mechanical system consists of an evaluation unit for the turntable and two trolleys positioned respectively on the left and right hand side of the turntable opposite to one another. The right trolley holds the detector-collimator system while the left one can be equipped with a transmission source (which has not been used here). Both trolleys can move horizontally towards and from the drum. The SGS is equipped with a high purity germanium detector (HPGe) connected to a cooling unit using liquid nitrogen. This detector is fixed in a lead collimator, enabling it to ``see" only a fraction of the waste package. The slice of a drum that can be seen by the detector is defined by the aperture of the collimator and the distance between detector and waste package. The output of the detector is amplified by a preamplifier and a spectroscopy linear amplifier. This pulse is converted into a digital number by the analogue-to-digital converter (ADC) for further computer-based processing, and is in our case combined with the multichannel analyzer (MCA). The used detector was calibrated by applying a custom correction function to another similar detector that is ISOCS/LabSOCS-calibrated \citep[][]{Canberra2020}. Data acquisition is performed and controlled by the main program which communicates with the Genie-2000 \citep[][]{Genie2000} software for conventional activity estimation.

Note that for nuclear fuel measurements, better activity estimates could be obtained by using a dedicated isotopic composition software \citep[e.g., FRAM, MGA][]{fram,mga} instead of Genie-2000. However, here estimation accuracy is not as critical as usual because these values only serve to define a weakly informative prior distribution for the MCMC inference (see section \ref{bayes}).

\subsection{Waste package and setup characteristics}
\label{setup}

We used the 3AX scanning results of a real non-conditioned radioactive waste package. The latter, a galvanized 200-liter drum (height: 890 mm; diameter: 580 mm; wall thickness: 1.25 mm), contains waste originating from the decommissioning of a MOX (Mixed Oxide Fuel) glove box. The matrix of the waste mainly consists of metals such as carbon steel ($\sim$70 wt\%) and lead ($\sim$8 wt\%). Additionally, a limited amount of mainly halogen-containing materials mixed with non-combustible waste ($\sim$22 wt\%) is present. The drum was filled up completely with a net weight of 54 kg. This roughly corresponds to an average matrix density of 0.27 g/cm$^{3}$. The drum dimensions, filling degree, matrix composition and matrix density distribution are considered fixed for this study and the associated uncertainties are therefore not taken into account. Instead, we focus on one of the main sources of uncertainties for this kind of measurements: the source distribution uncertainty (see section \ref{eff_model} for details).

\begin{figure}[hbt!]
	\noindent\hspace{1cm}\includegraphics[width=35pc]{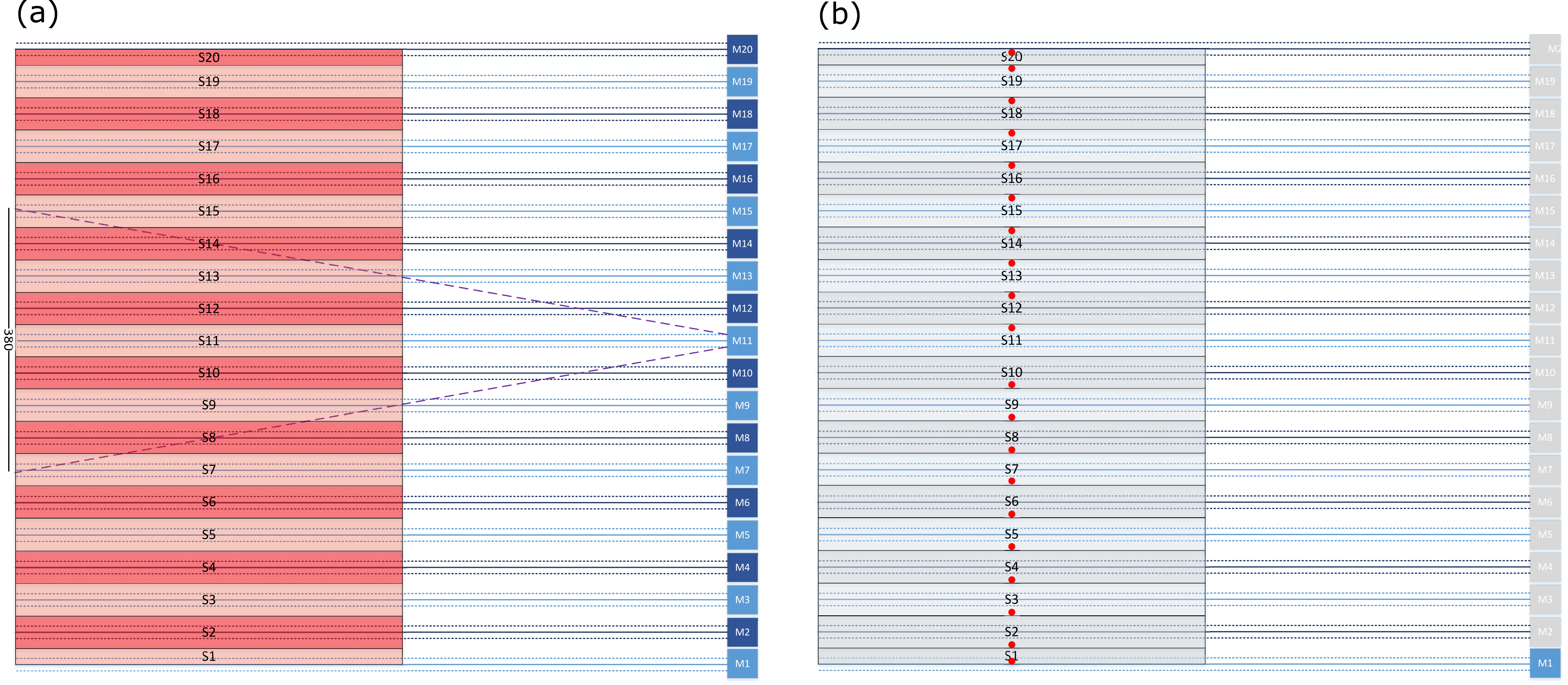}
	\caption{Vertical cross sections of the 3AX measuring setup for a 200-L drum. (a) 20 positions of the detector (M1 up to M20, blue) indicating the solid angle umbra (blue) for each measurement and the penumbra for measurement M11. The contents of the 200-L drum is discretized into 20 cylindrical volumes (segments) (red): S2 up to S19 in the middle of the drum with a height of 46.8 mm and S1 (bottom) \& S20 (top) with a height of 23.4 mm. (b) Homogeneous (gray rectangles) and point  (red dots) source distribution within each segment.}
	\label{fig1}
\end{figure}

Considering 20 individual measurements starting from the bottom of the drum and raising the detector 46.8 mm for each next measurement, the drum can be discretized into 20 horizontal segments (Figure \ref{fig1}): two segments of 23.4 mm height for bottom and top measurement (S1 \& S20) and 18 segments of 46.8 mm height (S2 $-$ S19). Measurement setup details are as follows:
\begin{itemize}
	\item The detector is surrounded by a rectangular lead slit collimator with a 1 mm copper coating (5$^{\circ}$ opening; 18 mm x 110 mm; 100 mm depth);
	\item The 200-liter drum is placed on a turning table and is continuously rotated during the measurements, with a rotation rate of 10 rotations per minute.
	\item The measured 20 segments result in 20 individual spectra (M1 up to M20) and measurement time is 300 s. For the first measurement, the bottom of the drum is placed in front of the middle of the detector (M1). For each following measurement the detector is raised by 46.8 mm. In this way, the top of the drum is placed in front of the middle of the detector for the 20$^{\rm th}$ measurement (M20, see Figure \ref{fig1}). 
\end{itemize}

\FloatBarrier

Table \ref{table1} lists the identified nuclides in the drum and their associated energy peaks.
\begin{table}[hbt!]
	\caption{Detected nuclides and associated photon energies.}
	\begin{adjustbox}{center}
		\begin{tabular}{cc}%
			\hline
			Nuclide & Energy [keV] \\
			\hline
			\multirow{3}{*}{Am-241} & 125.3 \\
			& 662.4 \\
			& 722.01\\
			& \\
			\multirow{2}{*}{Pu-238} & 152.72 \\
			& 766.39 \\
			& \\
			\multirow{5}{*}{Pu-239} & 129.30 \\
			& 345.01 \\
			& 375.05 \\
			& 413.71 \\
			& 451.48 \\
			& \\
			Pu-240 & 160.31 \\
			& \\
			Pu-241 & 148.57 \\
			\hline
		\end{tabular}
	\end{adjustbox}
	\label{table1}
\end{table}
\FloatBarrier

\subsection{Efficiency model}
\label{eff_model}

To account for the effect of the uncertainty associated with the source distribution within the drum on the detector efficiencies, we modeled the whole measurement system using the complex cylinder model of the Geometry Composer V4.3 library of the ISOCS/LabSOCS software by \citet{Canberra2020}. Taking the drum rotation during the measurement into account, we modeled two extreme source distributions (Figure \ref{fig1}):
\begin{itemize}
	\item Optimistic efficiency hypothesis: a homogeneous source distribution (hypothesis $h$) within each segment (Figure \ref{fig1}a). This is the maximum efficiency for the considered fixed matrix properties.
	\item Conservative efficiency hypothesis: a point source (hypothesis $p$) placed on the drum axes on the border of a segment (on the bottom of the segment for S1 up to S10 and on the top of the segment for S11 up to S20, see Figure \ref{fig1}b). This is the minimum efficiency for the considered fixed matrix properties.
\end{itemize}
Drum dimensions, filling degree and mean matrix composition and density are kept fixed in the ISOCS model and the only parameter that is varied is the source distribution. In case of the homogeneous source distribution, the matrix shielding is minimal for the part of the source close to the detector and increases more towards the center of the drum. For the point source on the drum axes, the matrix shielding thickness is larger than the drum radius. Since there is symmetry between measurements M1-M10 and measurements M11-M20 (Figure \ref{fig2}), the efficiency calculations for the lower part of the drum could be duplicated to the upper part of the drum. We designed an individual model for every individual measurement at the lower part of the drum (M1 up to M10), placing the source stepwise in segments S1 up to S20 for the homogeneous (hypothesis $h$) and point source (hypothesis $p$) scenarios, respectively. This means that we designed a total of 400 models (10 detector locations $\times$ 20 source locations $\times$ 2 source distributions). Overall, for both the $h$ and $p$ assumptions, the detector's efficiency was thus calculated for each configuration of source location (S, the source is located in a given segment while the other 19 segments do not contain any source), detector location (M), and photon energy (among the 12 energies listed in Table \ref{table1}). These efficiencies were then encapsulated into 2 $n_{source}$ $\times$ $n_{peaks}$ $\times$ $n_{seg}$ 3D arrays, $\textbf{E}^h$ and $\textbf{E}^p$, for the $h$ and $p$ assumptions, respectively. Here $n_{source} = n_{seg} = 20$ and $n_{peaks} = 12$.

\subsection{Count simulation model}
\label{count_model}

By using a $\lambda \in \left[0,1\right]$ coefficient, one can balance the efficiency $\epsilon\left(e\right)$ associated with a given detection at energy $e$ between the $h$ and $p$ assumptions for a given segment : $\epsilon\left(e\right) = \lambda\epsilon^h\left(e\right) + \left(1 - \lambda\right)\epsilon^p\left(e\right)$. This gives rise to our following count simulation model for a given count rate, $c^{rate}$

\begin{equation}
	\label{eq1}
	c^{rate}_{i,k} = \sum\limits_{l=1}^{n_{seg}}a_{l}^{j}P_{\gamma_{j,k}}\left[\lambda_l\epsilon^h_{i,l}\left(e_k\right)+\left(1 - \lambda_l\right)\epsilon^p_{i,l}\left(e_k\right)\right],
\end{equation}

where the subscript $i$ indexes the detector location, $i=1, \cdots, 20$, the subscript $k$ indexes the considered energy peak,  $k=1,\cdots,12$, the superscript $j$ indexes the considered nuclide, $j=1, \cdots, 5$, that is responsible for peak $k$, $a_{l}^{j}$ denotes the activity of nuclide $j$ at segment location $l$, and $P_{\gamma_{j,k}}$ is the emission probability of nuclide $j$ at energy $k$.

The simulated gross count at detector location $i$ for the energy $k$ is then calculated as

\begin{equation}
	\label{eq2}
	c^{gross}_{i,k} = c^{rate}_{i,k}t_m+b_{i,k},
\end{equation}

where $t_m$ is the measurement time and $b_{i,k}$ is the background continuum count for time $t_m$ at detector location $i$ for the energy $k$.

\subsection{Bayesian inference}
\label{bayes}

A common representation of the forward problem is
\begin{equation}
\textbf{d} = F\left(\bm{\uptheta}\right) + \textbf{e},
\label{mcmc0}
\end{equation}
where $\textbf{d} = \left[d_1, \ldots, d_{N_d} \right] \in \mathbb{R}^{N_d}, N_d \geq 1$ is the measurement data, $F\left(\bm{\uptheta}\right)$ is a deterministic forward model with parameters $\bm{\uptheta}$ and the noise term $\textbf{e}$ lumps all sources of errors. 

In the Bayesian paradigm, parameters in $\bm{\uptheta}$ are viewed as random variables with a posterior probability density function (pdf), $p\left(\bm{\uptheta} | \textbf{d} \right)$, given by

\begin{equation}
	p\left(\bm{\uptheta} | \textbf{d}  \right) = \frac{p \left(\bm{\uptheta}\right) p \left(\textbf{d} | \bm{\uptheta}\right)}{p \left( \textbf{d} \right)} \propto p\left(\bm{\uptheta}\right) L\left(\bm{\uptheta} | \textbf{d}\right),
	\label{mcmc1}
\end{equation}

where $L\left(\bm{\uptheta} | \textbf{d}\right) \equiv p \left(\textbf{d} | \bm{\uptheta}\right)$ signifies the likelihood function of $\bm{\uptheta}$. The normalization factor $p \left( \textbf{d} \right) = \int  p\left(\bm{\uptheta}\right) p\left(\textbf{d} | \bm{\uptheta}\right) d\bm{\uptheta}$ is not required for parameter inference when the parameter dimensionality is fixed. In the remainder of this paper, we will thus focus on the unnormalized density $p\left(\bm{\uptheta} | \textbf{d} \right) \propto p\left(\bm{\uptheta}\right) L\left(\bm{\uptheta} | \textbf{d}\right)$. 

If we assume $d$ to follow a Poisson process, which is the norm for count data, $L \left(\bm{\uptheta} | \textbf{d}\right)$ can be written as

\begin{equation}
	L\left(\bm{\uptheta} | \textbf{d}\right) = \prod\limits_{i=1}^{N_d}\exp\left(-\tilde{d_i}\right)\displaystyle\frac{\tilde{d_i}^d}{d!},
	\label{mcmc2}
\end{equation}

where $\tilde{\textbf{d}} = \left[\tilde{d}_1, \ldots, \tilde{d}_{N_d} \right] = F\left(\bm{\uptheta}\right)$ contains the simulated responses.

Here the vector of inferred variables, $\bm{\uptheta}$, consists of the 100 activities, $\textbf{a}$ (5 nuclides $\times$ 20 segments), 240 background continuum counts $\textbf{b}$ (12 energy peaks $\times$  20 segments) and 20 $\lambda$ coefficients. As of the $\textbf{d}$ vector, it comprises $N_{d} = 240$ gross counts obtained from the sum of the 240 net counts and 240 background continuum counts. In addition, our derived posterior distribution is not only conditioned to $\textbf{d}$ but also to the chosen $\textbf{E}^h$ and $\textbf{E}^p$ arrays. This results in assessment of the $p\left(\textbf{a},\textbf{b},  \bm{\uplambda} | \textbf{d}, \textbf{E}^h,\textbf{E}^p\right)$ distribution

\begin{equation}
	p\left(\textbf{a},\textbf{b},  \bm{\uplambda} | \textbf{d}, \textbf{E}^h,\textbf{E}^p\right) \propto L\left(\textbf{a},\textbf{b},  \bm{\uplambda} | \textbf{d},\textbf{E}^h,\textbf{E}^p\right)p\left(\textbf{a},\textbf{b},  \bm{\uplambda}\right).
	\label{mcmc3}
\end{equation}

The marginal posterior distribution of a given quantity, say $a_1$, is obtained by integrating the posterior distribution over all other inferred variables

\begin{equation}
	p\left(a_1 | \textbf{d}, \textbf{E}^h,\textbf{E}^p\right) = \int \int \int p\left(\textbf{a}_{\sim 1},\textbf{b},  \bm{\uplambda} | \textbf{d}, \textbf{E}^h,\textbf{E}^p\right)d\textbf{a}_{\sim 1} d\textbf{b} d\bm{\uplambda},
	\label{mcmc4}
\end{equation}

where the $\textbf{a}_{\sim 1}$ vector contain all elements of $\textbf{a}$ but $a_1$.

We assume the prior distributions for $\textbf{a}$, $\textbf{b}$ and $\bm{\uplambda}$ to be independent, which gives

\begin{equation}
	p\left(\textbf{a},\textbf{b},  \bm{\uplambda}\right) = p\left(\textbf{a}\right)p\left(\textbf{b}\right)p\left(\bm{\uplambda}\right).
	\label{mcmc5}
\end{equation}

For the activities, $\textbf{a}$, we have some prior knowledge in the form of estimates, $\hat{\textbf{a}}$, derived by application of the Genie-2000 \citep[][]{Genie2000} procedure to each segment separately. Assuming that these estimates have an accuracy of $\displaystyle \pm$ 1 order of magnitude gives an accuracy in base 10 logarithmic scale of: $\log_{10}\left(\hat{\textbf{a}}\right) \displaystyle \pm 1$. Further assuming that $\log_{10}\left(\textbf{a}\right)$ is normally distributed and that $\log_{10}\left(\hat{\textbf{a}}\right) \displaystyle \pm 1$ provides a 99\% uncertainty interval induces a standard deviation, $\sigma_a$, for $\log\left(\textbf{a}\right)$  of $\sigma_a = \displaystyle\frac{1}{3}\log\left(10\right)$ where the $\log\left(10\right)$ term accounts for the change of basis between $\log_{10}\left(\cdot\right)$ and $\log\left(\cdot\right)$. This forms the rationale for using a lognormal prior for $\textbf{a}$:  
$p\left(\log\left(\textbf{a}\right)\right) = N\left(\log\left(\hat{\textbf{a}}\right),\sigma_a^2\textbf{I}\right)$. As demonstrated in section \ref{results} this is a weakly informative prior from which the posterior distribution will easily depart if needed to appropriately fit the count data. Note that the values in $\hat{\textbf{a}}$ are likely to be overestimations as interpreting each segment separately implies that all the measured counts at a given detector location are assigned to the activity of the corresponding segment, which is incorrect.

Cases for which the detected net count(s) for a given (set of) measurement(s) $\left\{i,k=1,\cdots,n_k\right\}$, are too small compared to the derived background continuum count(s) for the Genie-2000 procedure to estimate an activity value, are easily included in the analysis as for non-influential activity value(s), the Bayesian approach will simply return the prior distribution. Overall, these ``non-detects" concerned 43 out of the 100 inferred activities, including the 20 Pu-240 activities. In these cases the Genie-2000 procedure returns a calculated MDA which means that in absence of any other information, the ``true" value could be anything between zero and something somewhat close to the calculated MDA \citep[see][]{Curie1968, Knoll2010}. For those activities,  $\textbf{a}_{mda}$, we decided to also use a lognormal prior but with mean parameter set to MDA/2: $p\left(\log\left(\textbf{a}_{mda}\right)\right) = N\left(\log\left(0.5\textbf{MDA}\right),\sigma_a^2\textbf{I}\right)$. Indeed, some practitioners use half the MDAs as working values \citep[e.g.,][]{Kim2020}. This is arguably subjective and another choice could be to use an uniform prior between zero and some upper bound related to the MDA, say: $p\left(\textbf{a}_{mda}\right) = U\left(\textbf{0},2\textbf{MDA}\right)$. Note however that no matter the specified upper bound, the inference should point out automatically that activities above the MDA are inconsistent with the observed counts. This is further discussed in section \ref{discussion}.

Regarding the 240-dimensional background continuum count vector, $\textbf{b}$, the obvious choice is to set $p\left(\textbf{b}\right)$ as the product of independent Poisson prior distributions with shape parameter equal to the (rounded) measured background continuum counts: $Pois\left(\textbf{b}\right)$. However, for technical reasons linked to the used software (see next section) we cannot use a Poisson prior (in short, Poisson priors cannot be assigned to real variables). Instead we use for $p\left(\textbf{b}\right)$ an uncorrelated and independent normal prior distribution with mean and variance vectors both equal to the measured count values: $p\left(\textbf{b}\right) = N\left(\textbf{b},\textbf{C}_b\right)$ with $\textbf{C}_b$ a diagonal matrix with the $b_1,\cdots,b_{240}$ values as diagonal elements. A $N\left(x,x\right)$ distribution provides an increasingly accurate approximation to  $Pois\left(x\right)$ as $x$ increases \citep[see, e.g.,][]{Barbour1992}. The approximation is commonly deemed excellent for $x > 1000$ and reasonably accurate for $x > 10$ \citep[][]{SOCR-UCLA}. Lastly, the prior distribution for the 20-dimensional $\bm{\uplambda}$ vector is taken as a bounded uniform distribution: $U\left(\textbf{0},\textbf{1}\right)$.

As no analytical solution of the 360-dimensional $p\left(\textbf{a},\textbf{b},  \bm{\uplambda} | \textbf{d}, \textbf{E}^h,\textbf{E}^p\right)$ distribution is available, we sample from $p\left(\textbf{a},\textbf{b},  \bm{\uplambda} | \textbf{d}, \textbf{E}^h,\textbf{E}^p\right)$ by MCMC simulation \citep[see][]{Gelman2014} using a state-of-the-art implementation of the HMC sampler \citep[see][for an extensive description]{Neal2011,Betancourt2018}. Convergence of the MCMC to the posterior target is monitored by means of the potential scale reduction factor, $\hat{R}$ \citep{Gelman-Rubin1992,Gelman2014}, using four independent Markov chains evolved in parallel (see next section). The $\hat{R}$ statistic compares for each parameter of interest the average within-chain variance to the variance of all the Markov chains mixed together. The closer the values of these two variances, the closer to one the value of the $\hat{R}$ diagnostic. Values of $\hat{R}$ jointly smaller than 1.02 for all sampled variables are deemed to indicate convergence to a limiting distribution.

\subsection{Software implementation}
\label{software}

We used the open-source greta package \citep[][]{Golding2019} to perform the HMC-based MCMC sampling.The greta package is an R \citep{Rsoftware} interface to some of the MCMC sampling algorithms implemented in the Tensorflow-probability package \citep[TFP,][]{Dillon2017} which itself relies on the Tensorflow (TF) machine learning platform \citep[][]{tensorflow2016}. The standard language to interact with TFP is Python \citep{Python} and the TFP-Python syntax is somewhat complicated. The main advantage of greta is to make it straightforward to build probabilistic models by constructing a directed acyclic graph and to perform MCMC sampling with TFP, using a seamless R-like syntax. The most useful MCMC sampler available through greta and used herein is HMC \citep[][]{Neal2011}. The HMC sampler has proven to be quite efficient when the gradient of the likelihood or objective function can be calculated by applying the chain rule, using an automatic differentiation (AD) technique such as implemented in TFP. Not all numerical models are suited to AD (for instance, numerical solvers of partial differential equations are not) but numerical models used for routine interpretation of radiological characterization measurements typically are. The TFP MCMC algorithms can be run in parallel on both CPUs and GPUs which makes greta to be quite computationally efficient. In this study, we ran four separate HMC trials in parallel over 4 CPUs. Additionally, most of the pre- and post-processing was performed with the tidyverse collection of packages \citep{tidyverse}, together with a few other specific packages \citep{coda,ggdist,patchwork}.

\section{Results}
\label{results}

The four parallel HMC runs were performed for a total of 30,000 warmup iterations each, after which 10,000 posterior samples were collected per run. The $\hat{R}$ convergence criterion computed using these 4 Markov chains was below 1.02 for all of the 360 sampled variables, indicating official convergence to the posterior target by the four runs. The collected 40,000 samples were thus used to approximate $p\left(\textbf{a},\textbf{b},  \bm{\uplambda} | \textbf{d}, \textbf{E}^h,\textbf{E}^p\right)$. On the used 4-core notebook (equipped with Intel i7-6820HQ CPU @ 2.70GHz), achieving these 40,000 iterations per/run incurred a computational time of about 2 hours in total.

\subsection{Synthetic example}
\label{synth_res}

Before proceeding with the inversion of the real data, we study in this section whether our proposed approach is able to correctly recover the activities for a synthetic problem  for which the true values are known. We used the same drum, efficiency data and count simulation model as for the real case, but considered only one nuclide and energy peak: Pu-239 at 413.71 keV. The ``true" $\lambda$ was set to 1 in every segment except for segment 16 where it was set to 0. The Pu-239 activity was set to 1$\times 10^{8}$ Bq in every segment except for segments 5 and 16 were it was set to 1$\times 10^{11}$ Bq. The resulting $i=1,\cdots,20$ simulated net counts, $c^{net}_{true,i}$, were used as mean parameters of 20 Poisson distributions from which the 20 observed nets counts were drawn: $c^{net}_{obs,i} \propto Pois\left(c^{net}_{true,i}\right)$ (see Table \ref{table2}). The 20 observed background continuum counts, $b_{obs,i}$, were also sampled from a Poisson distribution with a mean parameter inspired from the real measurements: $c^{back}_{obs,i} \propto Pois\left(100\right)$ (Table \ref{table2}). Note that due to the added errors in the observed net counts, no perfect inverse solution exists which is required for Bayesian inference (otherwise the posterior distribution would collapse to a single value and the MCMC sampling would inevitably fail to converge). The same normal $p\left(\textbf{b}\right)$ prior distribution and uniform $p\left(\bm{\uplambda}\right)$ prior distribution as for the real case were used. We defined $p\left(\textbf{a}\right) = N\left(\log\left(\hat{\textbf{a}}_{synth}\right),3\sigma_a^2\textbf{I}\right)$ with the 20 elements of the $\hat{\textbf{a}}_{synth}$ vector being $1 \times 10^{9}$ Bq. This way, the prior mean of the Pu-239 log-activity is (i) one order of magnitude too large for the 18 segments where the ``true" Pu-239 activity of the waste ($1 \times 10^{8}$ Bq) is small compared to the background activity, and (ii) two order of magnitude too low for the 2 segments where the ``true" Pu-239 activity is large ($1 \times 10^{11}$ Bq). In addition, setting the standard deviation of the log-prior activity to $3\sigma_a^2$ (that is, $1\log\left(10\right)$) induces a wide and thus quite uncertain prior.

\begin{figure}[hbt!]
	\noindent\hspace{0cm}\includegraphics[width=40pc]{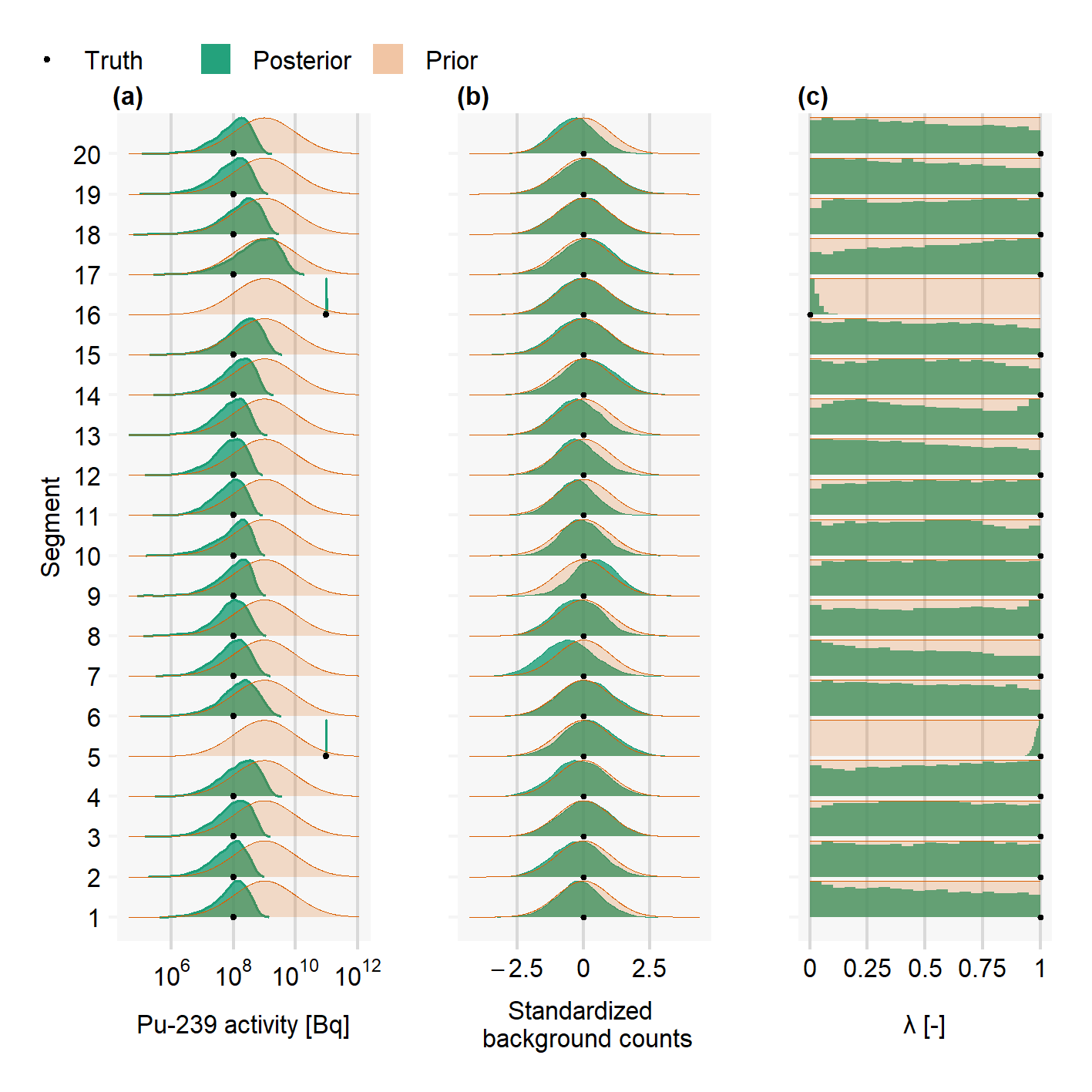}
	\caption{Prior and posterior (a) activity, (b) background continuum count and (c) $\lambda$ distributions for the synthetic test case. The black dots denote the true values used to generate the gross count data. Note that the $x$-axis of subplot (a) has a base 10 logarithmic scale. For visual convenience, the background continuum counts have been standardized using the normal prior parameters. For the $\lambda$ variable, we show a posterior histogram instead of a kernel density estimate. This is because when applied to a bounded data sample, kernel density smoothing tends to create artifacts near the bounds.} 
	\label{fig2}
\end{figure}

Figure \ref{fig2} presents the resulting posterior parameter distribution. Note that segment numbering goes from bottom (segment 1) to top (segment 20). With respect to the activities (Figure \ref{fig2}a), it is seen that the MCMC sampling recovers the 2 large true activities in segments 5 and 16 very accurately, with a high degree of certainty. This even though the mode of the lognormal prior is way off. The posterior log-activity distributions in the other segments (all but 5 and 16) often peak near the true values and are somewhat asymetric with a larger probability mass at the right side. This is because here the MCMC sampling finds a balance between simulating sufficiently small gross counts and honoring the prior. This nicely illustrates how Bayesian analysis handles MDAs (non-detects). If the range of values spanned by the prior distribution is sufficiently low, then it simply returns the prior distribution. Otherwise, the posterior distribution forms an intermediate distribution between the prior and the likelihood (equation (\ref{mcmc1})). The posterior background continuum count distributions globally peak around the true values, with slight deviations from their associated priors (Figure \ref{fig2}b). As expected, the posterior $\lambda$ distributions are the same as their prior counterparts for the low activities (Figure \ref{fig2}c). For the 2 segments where the true activity is large, the inference put most of the probability mass towards the true boundary values of $\lambda$, $\lambda = 1$ for segment 5 and $\lambda = 0$ for segment 16. Overall these results demonstrate that our approach can correctly infer the joint posterior distribution of $\textbf{a}$, $\textbf{b}$ and $\bm{\uplambda}$.

\begin{table}[hbt!]
	\caption{Observed background continuum and net counts used in the synthetic experiment for each detector location (segment).}
	\begin{adjustbox}{center}
		\begin{tabular}{ccc}%
			\hline
			Detector location & Background counts & Net counts \\
			\hline
			1 & 102 & 18 \\
			2 & 96 & 188 \\
			3 & 97 & 1284 \\
			4 & 98 & 3844 \\
			5 & 101 & 6089 \\
			6 & 85 & 3953 \\
			7 & 98 & 1152 \\
			8 & 118 & 176 \\
			9 & 102 & 38 \\
			10 & 94 & 18 \\
			11 & 118 & 10 \\
			12 & 103 & 14 \\
			13 & 107 & 27 \\
			14 & 84 & 747 \\
			15 & 99 & 2452 \\
			16 & 105 & 4352 \\
			17 & 107 & 4380 \\
			18 & 108 & 2469 \\
			19 & 95 & 724 \\
			20 & 101 & 21 \\
			\hline
		\end{tabular}
	\end{adjustbox}
	\label{table2}
\end{table}
\FloatBarrier

\FloatBarrier

\subsection{Real data}
\label{real_res}

\begin{figure}[hbt!]
	\noindent\hspace{-1.5cm}\includegraphics[width=45pc]{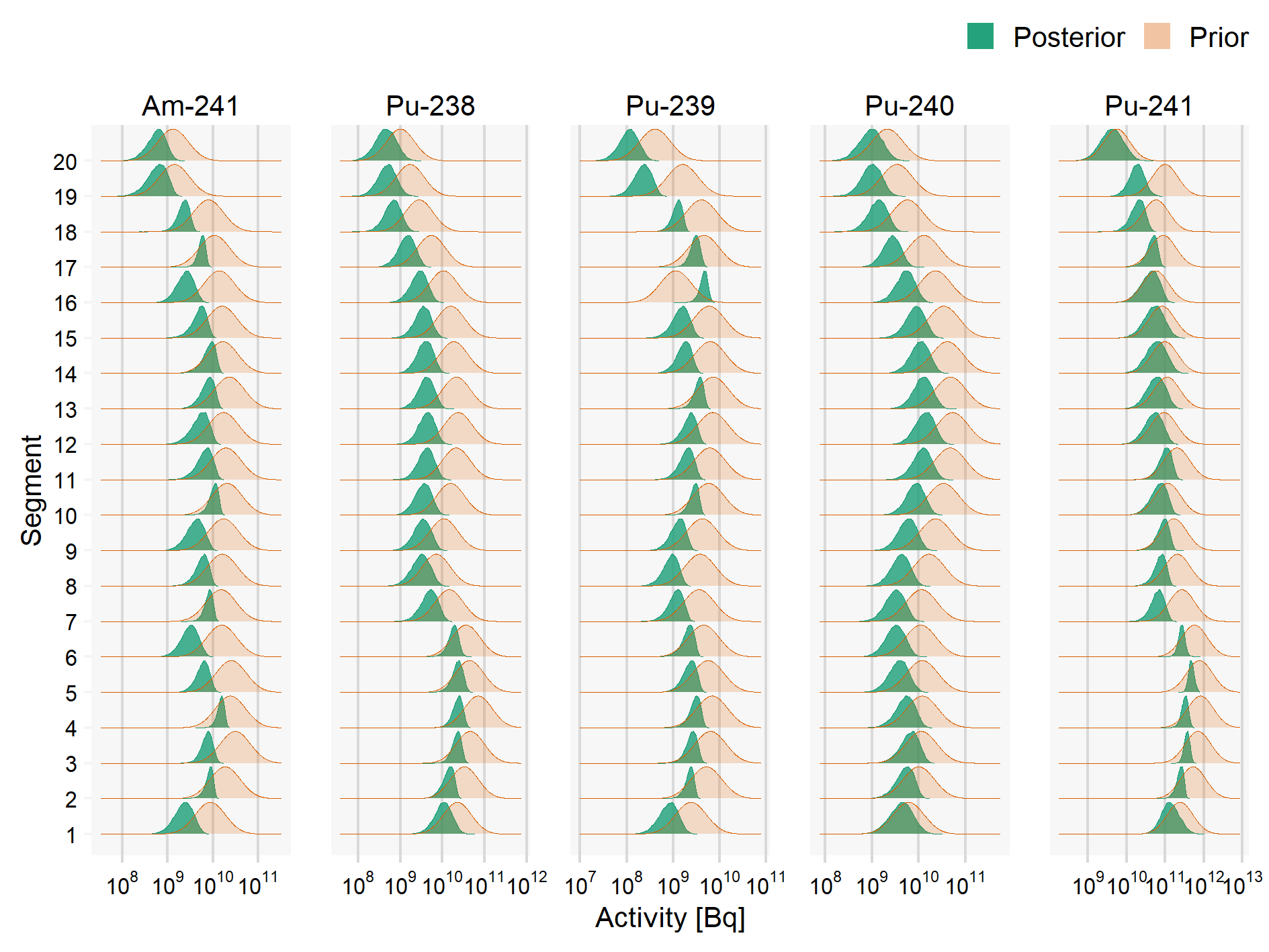}
	\caption{Prior and posterior distributions of the 100 considered activities. Segment numbering goes from bottom (1) to top (20). Note the base 10 logarithmic scale of the $x$-axis.} 
	\label{fig3}
\end{figure}

Figure \ref{fig3} depicts the marginal posterior distributions of the inferred Am-241, Pu-238, Pu-239, Pu-240 and Pu-241 activities: $p\left(\textbf{a}| \textbf{d}, \textbf{E}^h,\textbf{E}^p\right)$. It is observed that the ``multi-energy" nuclides, that is, Am-241, Pu-238 and Pu-239 which are associated with three (Am-241), two (Pu-238) and five (Pu-239) energy peaks, respectively (Table \ref{table1}), are globally well resolved. This is indicated by a posterior distribution that is narrower than the prior distribution. Strikingly, for all 100 activities but that of Pu-239 in segment 16, the posterior mode is lower than the prior mean of the log-activity ($\log\left(\hat{\textbf{a}}\right)$ or $\log\left(0.5\textbf{MDA}\right)$, see section \ref{bayes}). As written above, this is because processing each segment individually, as done to define $p\left(\textbf{a}\right)$, is likely to overestimate the actual activities as it overlooks the fact that the detected counts at a given detector location are not only caused by the activity in the considered segment but also by the activities in the neighboring segments. In contrast, for a given nuclide our net count rate simulation model (see equation (\ref{eq1}) and associated text) accounts for the contributions from all activities to each measurement. For Pu-239 at segment 16 only (Figure \ref{fig3}), the posterior mode of the log-activity is above the selected prior mean: about 3.9 $\times$ 10$^{9}$ Bq against 1.13 $\times$ 10$^{9}$ Bq. Note that since this activity is among those for which the Genie-2000 procedure returned a MDA value, here the posterior mode is about two times the MDA (prior mean of the log-activity was taken as $\log\left(MDA/2\right)$).

The ``single-energy" Pu-240 and Pu-241 nuclides are relatively well resolved though, perhaps not surprisingly, some are more uncertain than the multi-energy nuclides. This is especially the case for Pu-240 at segment 1 and Pu-241 at segments 13 $-$ 16 and 20. Overall, at these locations the 95\% uncertainty intervals for Pu-240 and Pu-241 cover about one order of magnitude.

Table \ref{table3} lists the corresponding 95\% posterior uncertainty (or credible) intervals of the total activity of each nuclide over the whole drum. For this package, the Pu-241 nuclide has by far the largest activity, one to two orders of magnitude higher than for the other nuclides. As written earlier, the measured net count for the 20 Pu-240 activities is zero and, in this case, the prior distribution was set to $N\left(\log\left(0.5\textbf{MDA}\right),\sigma_a^2\textbf{I}\right)$. Since the range spanned by the posterior Pu-240 log-activities is systematically lower than the range spanned by the prior Pu-240 log-activities, our results show that for the considered case study setting the Pu-240 non-detects to half the MDA is overly conservative.

\begin{table}[hbt!]
	\caption{95\% posterior uncertainty intervals [10$^{9}$ Bq] of the 5 nuclides' activities across the whole drum.}
	\begin{adjustbox}{center}
		\begin{tabular}{cc}%
			\hline
			Nuclide & 95\% uncertainty [10$^{9}$ Bq] \\
			\hline
			Am-241 & $112.17 - 118.83$ \\
			Pu-238 & $142.84 - 170.17$ \\
			Pu-239 & $38.15 - 40.26$ \\
			Pu-240 & $101.71 - 152.70$ \\
			Pu-241 & $2436.79 - 2882.08$ \\
			\hline
		\end{tabular}
	\end{adjustbox}
	\label{table3}
\end{table}

\FloatBarrier

\begin{figure}[hbt!]
	\noindent\hspace{0cm}\includegraphics[width=40pc]{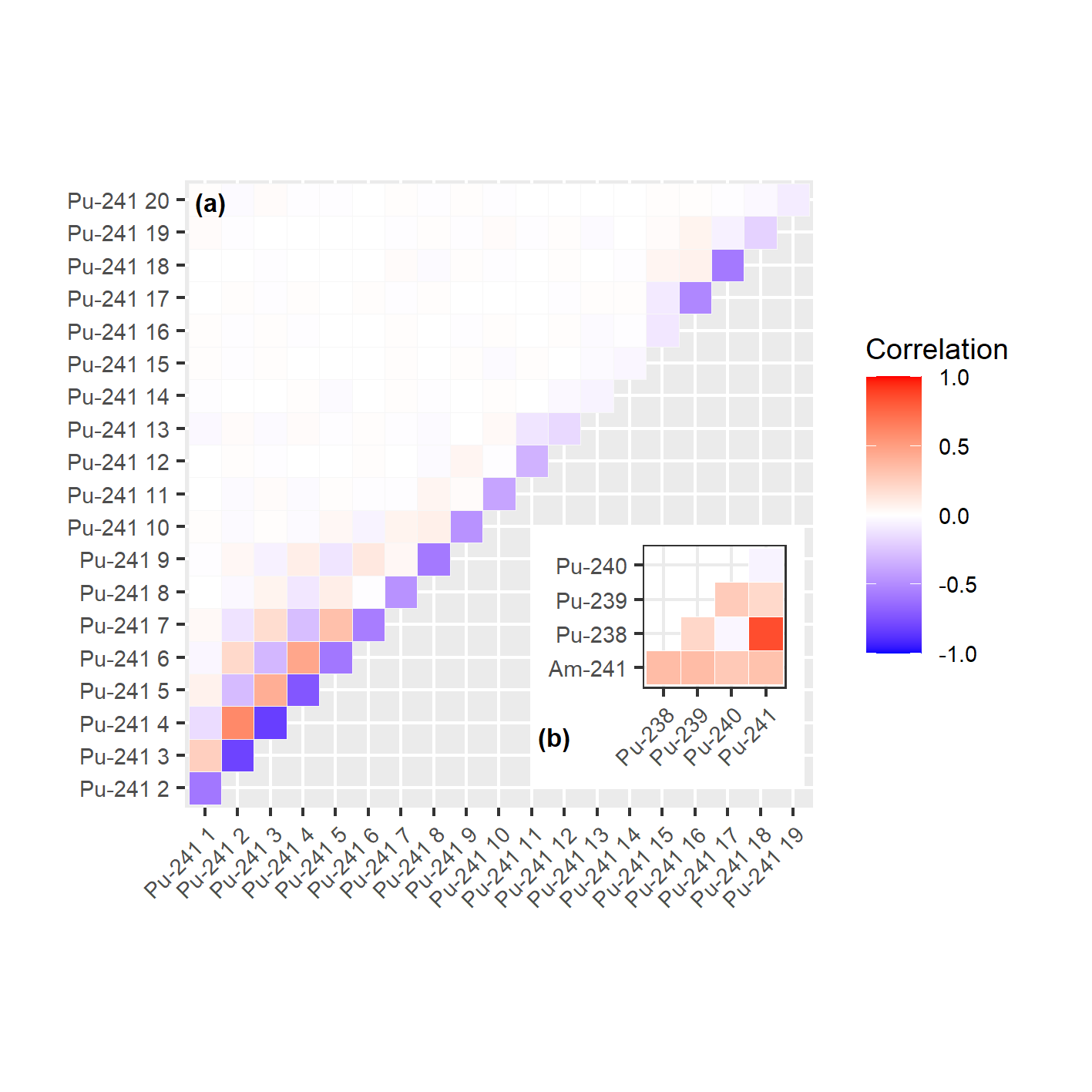}
	\caption{Posterior (Pearson) correlations corresponding to every pair of inferred activities.} 
	\label{fig4}
\end{figure}

Figure \ref{fig4} provides more insights into the posterior activity distribution by displaying the (Pearson) linear correlation coefficients between (a) the inferred Pu-241 activities in each segment (Figure \ref{fig4}a) and (b) the 5 nuclides across all segments (Figure \ref{fig4}b). The Pu-241 correlations between segments (Figure \ref{fig4}a) are representative of the between-segment correlations of the other nuclides. The following pattern is observed. For a given nuclide, there is a negative correlation between direct neighbor (or lag-1) activites. This makes sense as direct neighbors contribute more than segments further away to the simulated count at a considered location. Hence, the MCMC inference can counterbalance the effect of increasing the activity in a segment by decreasing the activity in a direct neighbor location. Lag-2 activities (that is, activities separated by one segment) then correspondingly show a weak positive correlation while correlations fade away from lag-3 on. Moreover, looking at the correlations between the 5 nuclides across all segments merged together (Figure \ref{fig4}b) reveals only one important correlation: the correlation coefficient between Pu-238 and Pu-241 is 0.85.

\begin{figure}[hbt!]
	\noindent\hspace{0cm}\includegraphics[width=35pc]{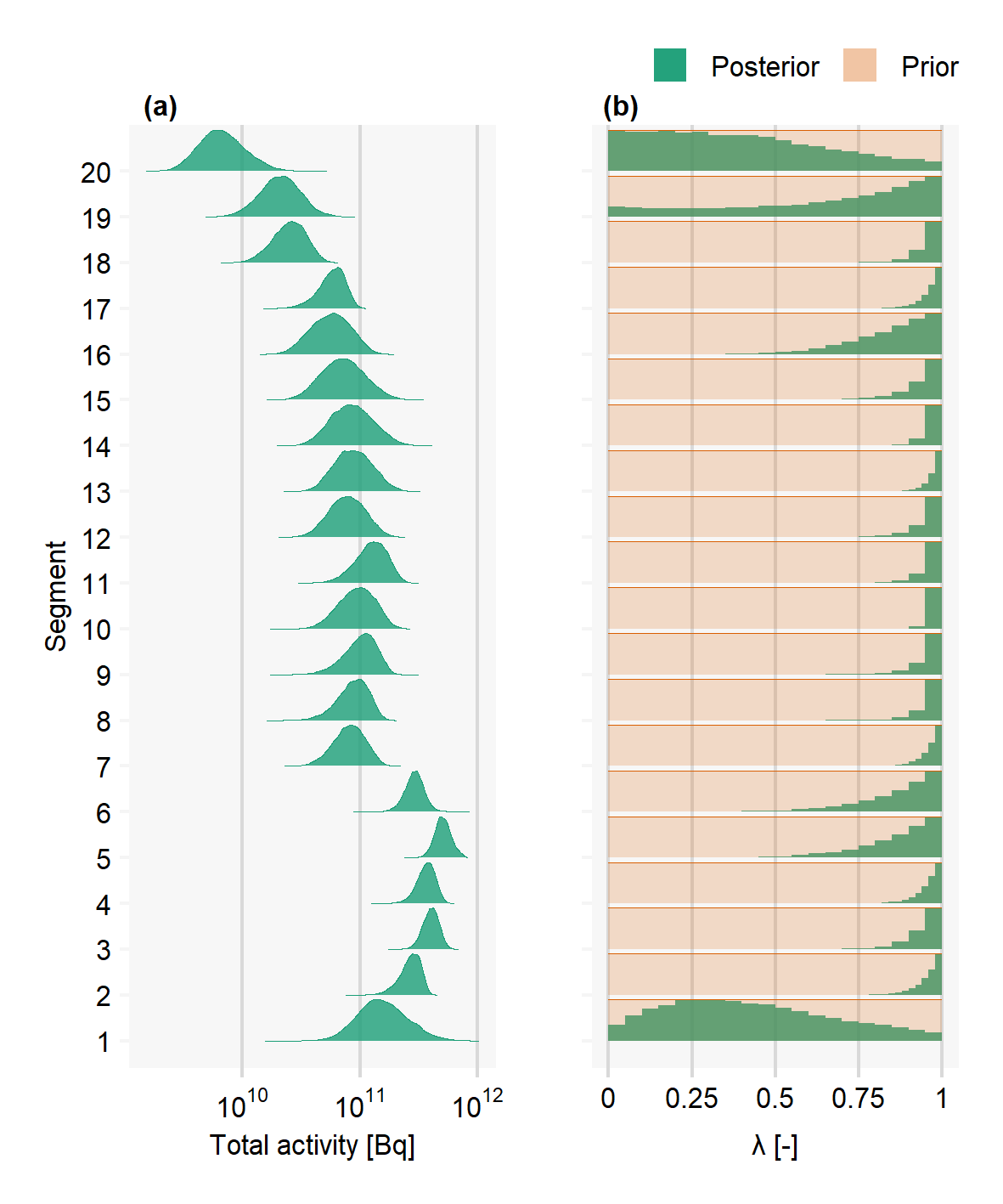}
	\caption{Posterior distributions of (a) the total activity and (b) the $\lambda$ parameter within each segment. The chosen uniform prior For $\lambda$ is also shown in subplot (b). No prior distribution is shown for the total activity as no closed form expression is available for the sum of lognormal distributions. Segment numbering goes from bottom (1) to top (20). Note that the $x$-axis of subplot (a) has a base 10 logarithmic scale. Note that for the $\lambda$ variable, we show a posterior histogram instead of a kernel density estimate. This is because when applied to a bounded data sample, kernel density smoothing tends to create artifacts near the bounds.} 
	\label{fig5}
\end{figure}

To complete this analysis of the posterior activity distribution, Figure \ref{fig5}a displays the total posterior activity through the considered drum, from bottom (segment 1) to top (segment 20). Globally, the total posterior activity decreases from bottom (segment 1) to top (segment 20), from about $1.5 \times 10^{11}$ Bq to less than $1 \times 10^{10}$ Bq.

\FloatBarrier

\begin{figure}[hbt!]
	\noindent\hspace{-1.5cm}\includegraphics[width=45pc]{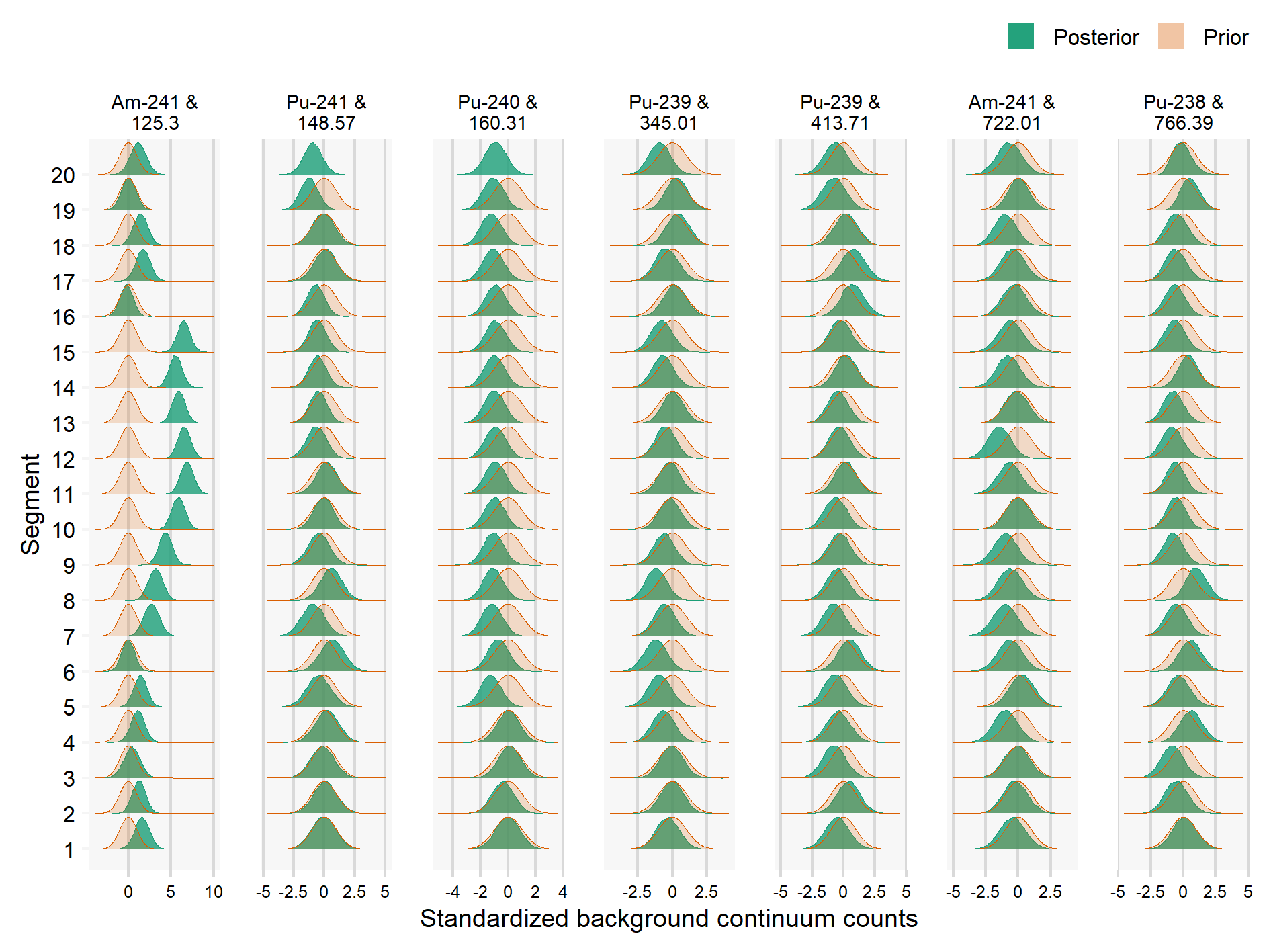}
	\caption{Prior and posterior distributions of the inferred background continuum counts. Each facet's label mentions the considered combination of nuclide and energy peak (keV). For visual convenience, the counts have been standardized using the normal prior parameters and only 7 out of the considered 12 peaks are shown (the following energy peaks are left out of the plot: Am-241 \& 662.24 keV, Pu-238 \& 152.72 keV, Pu-239 \& 129.3 keV, and Pu-239 \& 375.05 keV). Segment numbering goes from bottom (1) to top (20).} 
	\label{fig6}
\end{figure}

The marginal posterior background continuum activities, $p\left(\textbf{b}|\textbf{d}, \textbf{E}^h,\textbf{E}^p\right)$, are depicted in Figure \ref{fig6}. The posterior distributions are generally close to their associated prior distribution but some depart from it. Consistently with our Bayesian framework, these deviations are needed to maximize the posterior density. We would like to stress that the prior, $p\left(\textbf{b}\right)$, is based on a measured background continuum count that is considered to be a realization of a Poisson distribution of which the shape parameter is unknown. By setting $p\left(\textbf{b}\right) = Pois\left(\textbf{b}\right)$ (or $p\left(\textbf{b}\right) = N\left(\textbf{b},\textbf{C}_b\right) \approx Pois\left(\textbf{b}\right)$ as done herein) one assumes that the measured background continuum count is equal to the mean of its underlying distribution, which is obviously not necessarily the case. Some deviations from $p\left(\textbf{b}\right)$ should thus come at no surprise. If deemed necessary, an alternative solution to enforce a tight closeness between $N\left(\textbf{b},\textbf{C}_b\right) \approx Pois\left(\textbf{b}\right)$ and $p\left(\textbf{b}|\textbf{d}, \textbf{E}^h,\textbf{E}^p\right)$ is discussed in section \ref{discussion}.

\FloatBarrier

Figure \ref{fig5}b presents the marginal posterior densities of the elements of the $\bm{\uplambda}$ vector, which is our parameter that accounts for the uncertainty in the efficiencies. For segments 2 to 19 the posterior probability mass is mostly concentrated on the upper bound of 1, meaning that the MCMC is strongly favoring the hypothesis of a homogeneous radioactive source distribution in these locations. In contrast, for the bottom (1) and top (20) segments every intermediate state between a fully homogeneous source distribution and a fully point source distribution has non-zero probability. it is important to note that our efficiency model (see section \ref{eff_model}) might be biased, especially for the top and bottom segments. Indeed, a smaller filling height than the drum top may cause the large uncertainty in the source distribution for top segment 20 and the long-tail uncertainty distribution in the underlying segment 19. Furthermore, there is an empty space of up to 12 mm between the bottom of segment 1 and the ```floor" because the bottom plate of the drum does not touch the floor, only the drum's outer bottom ring does. This may thus be responsible for the large source distribution uncertainty for segment 1.

\begin{figure}[hbt!]
	\noindent\hspace{-0cm}\includegraphics[width=35pc]{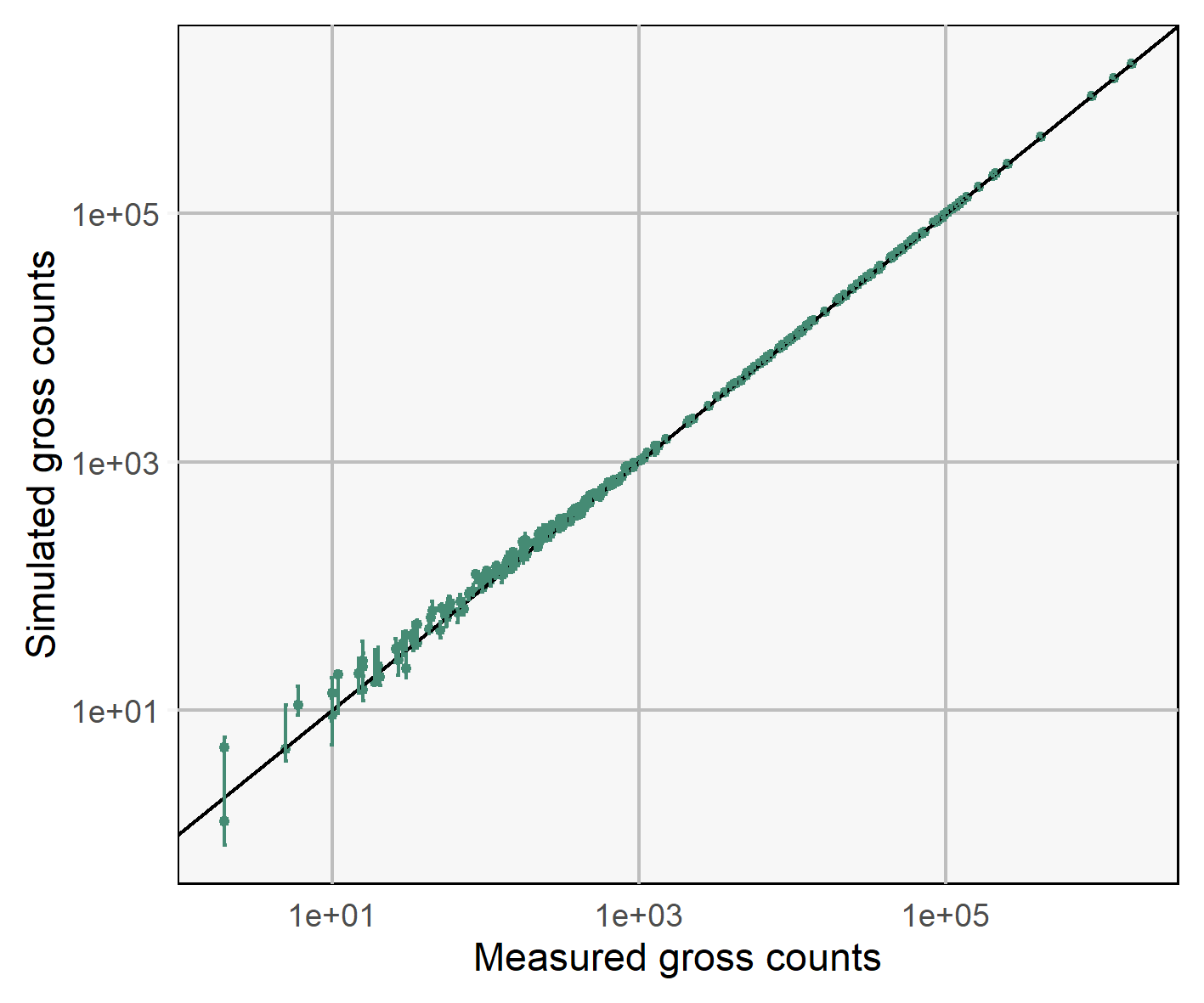}
	\caption{Measured gross counts against a posteriori simulated gross counts. The green points represent the maximum likelihood solution (solution with the largest Poisson likelihood) among the collected 40,000 posterior samples, and the vertical bars denote the 95\% posterior uncertainty intervals. A base 10 logarithmic scale is used to make the deviations between the smallest observed and simulated gross counts visible.} 
	\label{fig7}
\end{figure}

The fit of the posterior gross count simulations, $F\left(\textbf{a},\textbf{b},\bm{\uplambda},\textbf{E}^h,\textbf{E}^p\right)$ to the measured gross counts, $\textbf{d}$ is depicted in Figure \ref{fig7} using a base 10 logarithmic scale to make the discrepancies between (very) small measured and simulated counts visible. The vertical bars denote the 95\% posterior uncertainty intervals associated with the simulated counts. The observed counts are mostly well fitted and the 95\% uncertainty intervals are relatively tight and most of the time include the 1:1 line. Furthermore, the most important discrepancies, in relative terms, are for some rather small gross counts, below 10 $-$ 15. This is consistent with both our expectations and the Poisson count statistics.

\FloatBarrier

\section{Discussion}
\label{discussion}

We have shown that vertical activity profiles can be probabilistically inferred from SGS data accounting for the effect of the source distribution uncertainty on the detector efficiency. The following points and/or potential improvements nevertheless deserve further attention. 

\begin{figure}[hbt!]
	\noindent\hspace{-1.5cm}\includegraphics[width=45pc]{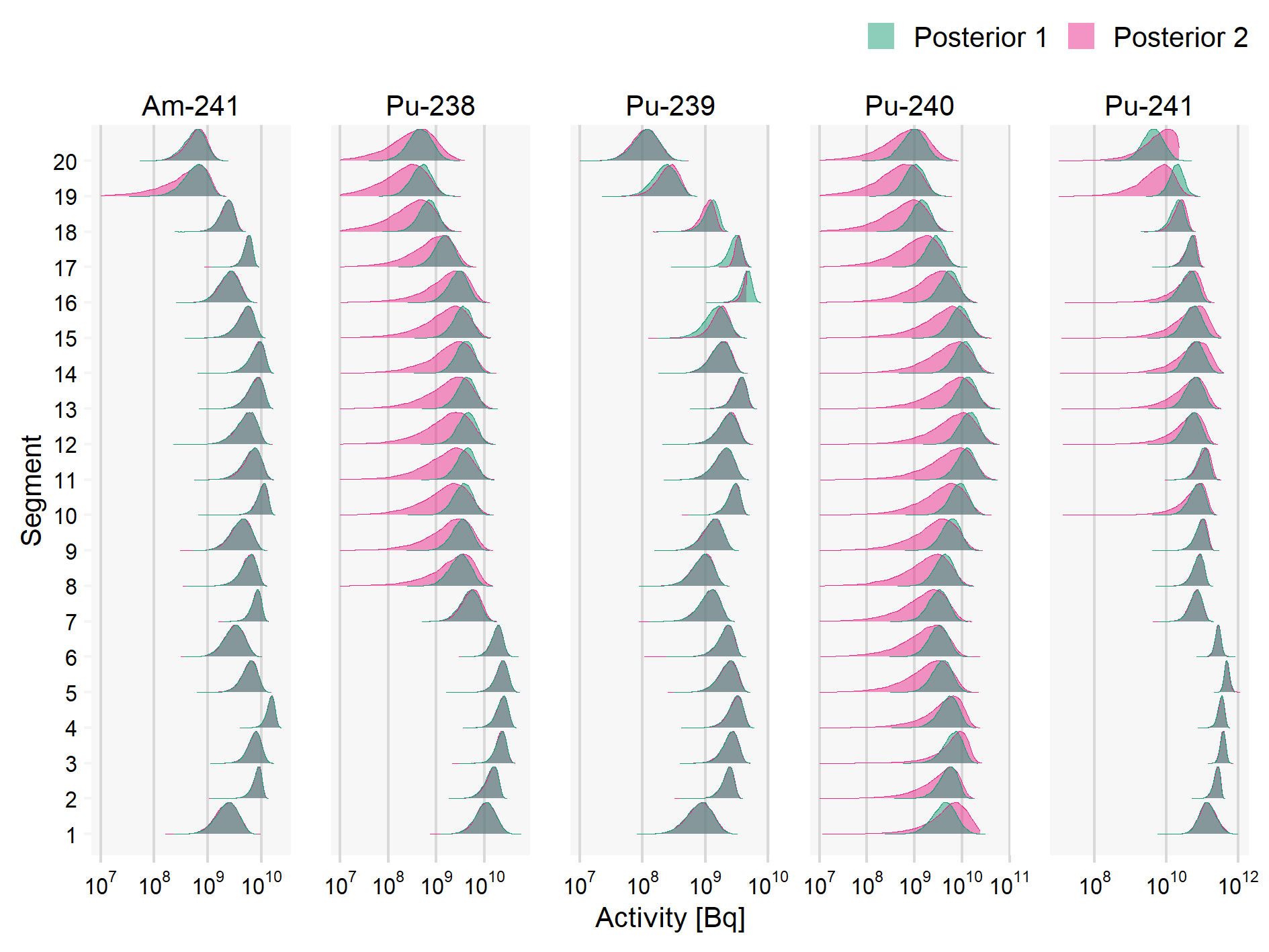}
	\caption{Posterior activity distributions obtained when using (1) $p\left(\log\left(\textbf{a}_{mda}\right)\right) = N\left(\log\left(0.5\textbf{MDA}\right),\sigma_a^2\textbf{I}\right)$ (Posterior 1) and (2) $p\left(\textbf{a}_{mda}\right) = U\left(\textbf{0},2\textbf{MDA}\right)$. For visual convenience, the lower limit of the $x$-axis is thresolded to $1 \times 10^{7}$ Bq.  Also, note the base 10 logarithmic scale of the $x$-axis.} 
	\label{fig8}
\end{figure}

As for any prior distribution, the choice of the priors for the activities flagged as MDAs by the standard analysis procedure \citep[][]{Genie2000} applied to each segment separately, $\textbf{a}_{mda}$, is a relatively subjective expert-based decision. Note, however, that if the prior is sufficiently weakly informative or wide and the information content of the count measurement data is large enough (sufficiently large net counts compared to the corresponding background counts and not too many missing values), then the actual shape of the prior should not substantially influence the posterior results. An additional MCMC run with using $p\left(\textbf{a}_{mda}\right) = U\left(\textbf{0},2\textbf{MDA}\right)$ instead of $p\left(\log\left(\textbf{a}_{mda}\right)\right) = N\left(\log\left(0.5\textbf{MDA}\right),\sigma_a^2\textbf{I}\right)$ shows that, unsurprisingly, $p\left(\textbf{a}_{mda}\right) = U\left(\textbf{0},2\textbf{MDA}\right)$ leads to posterior distributions for the elements of $\textbf{a}_{mda}$ that are close to their associated uniform prior (Figure \ref{fig8}). For the non-MDAs, the two posterior distributions are very similar (Figure \ref{fig8}).

In some relatively few cases, the marginal posterior background continuum activities, $p\left(\textbf{b}|\textbf{d}, \textbf{E}^h,\textbf{E}^p\right)$, differ largely from the used prior distribution for $\textbf{b}$, $N\left(\textbf{b},\textbf{C}_b\right) \approx Pois\left(\textbf{b}\right)$. This is probably due to some small inconsistencies between our model and the real drum. Notwithstanding, if having $p\left(\textbf{b}|\textbf{d}, \textbf{E}^h,\textbf{E}^p\right)$ closely approximating $N\left(\textbf{b},\textbf{C}_b\right) \approx Pois\left(\textbf{b}\right)$ is judged necessary, then setting $p\left(\textbf{b}\right) = N\left(\textbf{b},s_b\textbf{C}_b\right)$ with $s_b < 1$ can do the job. Indeed, we found for the considered case study that using $s_b = 0.01$ allows for $p\left(\textbf{b}|\textbf{d}, \textbf{E}^h,\textbf{E}^p\right)$ to closely reproduce $N\left(\textbf{b},\textbf{C}_b\right)$ while only slightly changing $p\left(\textbf{a}| \textbf{d}, \textbf{E}^h,\textbf{E}^p\right)$. However, this comes at the expense of a less good fit of the simulated gross counts, $F\left(\textbf{a},\textbf{b},\bm{\uplambda},\textbf{E}^h,\textbf{E}^p\right)$ to the measured gross counts, $\textbf{d}$ (not shown).

This study only considers the count measurement uncertainty and the uncertainty of the source distribution (albeit in a simplified way), the latter being one important source of uncertainty in the detector's efficiency. Nonetheless, the matrix density distribution of the package is assumed to be homogeneous. Since the matrix-related uncertainties can be relatively large and thereby impacting the detector's efficiency too, more work is needed to account for it within the inference. This could be approached by using a Monte Carlo particle transport code to simulate the impact of various matrix and source configurations on the efficiency of the considered detector. The so-derived dataset could then be used to construct a computationally cheap nonlinear regression model (also called proxy or metamodel) that would predict the simulated efficiencies from the sampled matrix and source parameters, for each considered nuclide and segment location. Both matrix-related and source distribution uncertainty could thus be accounted for within the Bayesian inference by sampling the surrogate model parameters along with the other variables. This warrants further investigations.

More generally, an extended Bayesian approach based on the one presented herein could also be used to infer spatial activity distributions in 1D, 2D and 3D within a drum waste package from combined measurements from different techniques. This will be investigated in future work, in the framework of the MICADO EU project (see section \ref{acknowledgments}).

\FloatBarrier

\section{Conclusion}
\label{conclusion}

We propose a Bayesian approach to probabilistically infer vertical activity distributions within a radioactive waste drum from segmented gamma scanning (SGS) measurements. Our approach relies on state-of-the-art Markov chain Monte Carlo (MCMC) sampling using the Hamiltonian Monte Carlo (HMC) technique and accounts for the net and background count measurement uncertainty together with the uncertainty in the detector efficiencies caused by the uncertainty in the source distribution within the drum. Furthermore, our efficiency model accounts for the contributions of all considered segments to each count measurement. For the considered case study, our approach is used to resolve the vertical activity distribution of 5 nuclides over 20 locations and produce sound uncertainty estimates. Main topics for future research include how to incorporate the impact of the uncertainty in matrix density distribution on the modeled efficiencies, to extend our setup from 1D to 2D and/or 3D, and to combine measurements from different techniques for better radionuclides' quantification.

\section{Acknowledgments}
\label{acknowledgments}

This work received funding by the EU project MICADO Consortium. 2018: ``Horizon 2020, Nfrp 2018-10, Proposal 847641: Measurement and Instrumentation for Cleaning and Decommissioning Operations".
\appendix




\end{document}